\documentclass[showpacs,reprint,flushbottom,linenumbers,aps,prl]{revtex4-1}

\usepackage{amsmath, amssymb, graphicx, psfrag, color}

\def\Gcm2{\mbox{ G$\times$cm$^2$}}

\def\Hz{\mbox{ Hz}}

\def\ea0{\mbox{ $ea_0$}}

\def\V{\mbox{ V}}

\def\kV{\mbox{ kV}}

\def\mF0{\mbox{ $\mu\Phi_0$}}
\def\mF0rtHz{\mbox{ $\mu\Phi_0/\sqrt{\rm Hz}$}}

\def\ecm{~e\mbox{cm}}

\def\uCcm{\mbox{ $\mu$C/cm$^2$}}

\usepackage{txfonts}

\graphicspath{{./}{./figs/}}

\begin{document}

\title{A limit on the electron electric dipole moment using paramagnetic ferroelectric Eu$_{0.5}$Ba$_{0.5}$TiO$_3$}

\author{S. Eckel}
\email{stephen.eckel@aya.yale.edu}
\altaffiliation{Present Address: Joint Quantum Institute, National Institute of Standards and Technology, Gaithersburg, MD 20899, USA}
\author{A.O. Sushkov}
\altaffiliation{Present Address: Physics Department, Harvard University, Cambridge, Massachusetts 02138, USA}
\author{S.K. Lamoreaux}
\affiliation{Yale University, P.O. Box 208120 New Haven, CT 06520-8120}


\date{\today}

\begin{abstract}
We report on the results of a search for the electron electric dipole moment $d_e$ using paramagnetic ferroelectric Eu$_{0.5}$Ba$_{0.5}$TiO$_3$.  The electric polarization creates an effective electric field that makes it energetically favorable for the spins of the seven unpaired $4f$ electrons of the Eu$^{2+}$ to orient along the polarization, provided that $d_e\neq 0$.  This interaction gives rise to sample magnetization, correlated with its electric polarization, and is therefore equivalent to a linear magnetoelectric effect.  A SQUID magnetometer is used to search for the resulting magnetization.  We obtain $d_e = (-1.07\pm3.06_\text{stat}\pm 1.74_\text{sys})\times10^{-25}\ecm$, implying an upper limit of $|d_e|<6.05\times10^{-25}\ecm$ (90\% confidence).
\end{abstract}

\maketitle

The permanent electron electric dipole moment ($e$EDM) has been of experimental interest for nearly half a century because it provides a probe of charge-parity (CP) symmetry violation in the universe.  Through the CPT theorem~\cite{Schwinger1951}, the existence of a permanent electric dipole moment, which violates time-reversal (T) symmetry, would imply violation of CP in order that combined operations of CPT are conserved.  CP symmetry violation is required in the early universe in order to explain the currently observed matter-antimatter asymmetry~\cite{Sakharov1967}; furthermore, the CP violation in the standard model (SM) is not sufficient to explain this asymmetry~\cite{Farrar1994}.  Many theories that go beyond the SM contain more CP violation and therefore predict a larger $e$EDM that may be detected by the next generation of experiments~\cite{Bernreuther1991}.

The traditional method to search for an $e$EDM involves observing precession of an atom or molecule with unpaired electron spins in the presence of both magnetic and electric fields~\cite{KhriplovichLamoreaux1997}.  This method has been used extensively~\cite{Regan2002,Griffith2009} and has set the best current upper limit on the $e$EDM of $|d_e| < 1.05\times10^{-27}\ e\mbox{cm}$~\cite{Hudson2011}.  Another measurement procedure, first suggested by Shapiro~\cite{Shapiro1968}, involves placing unpaired election spins bound to a crystal lattice in an electric field.  If $d_e\neq0$, the electrons will orient along the electric field and produce a magnetization in the sample~\cite{Budker2006}.  To date, two experiments produced $e$EDM limits using this approach~\cite{Vasilev1978,Kim2011}.  The reverse experiment, where the sample is magnetized and a correlated polarization is measured, has also been performed~\cite{Heidenreich2005}.  These solid-state-based experiments sacrifice the narrow atomic and molecular transition linewidths for a significantly larger signal due to the high density of spins present in a solid.

Perhaps the most important choice for a solid-state $e$EDM experiment is the material.  In Refs.~\cite{Rushchanskii2010,Sushkov2010}, the advantages of Eu$_{0.5}$Ba$_{0.5}$TiO$_3$ are detailed over other materials, and a short review will be presented here.  Eu$_{0.5}$Ba$_{0.5}$TiO$_3$ has a perovskite crystal structure and is ferroelectric below approximately 200~K~\cite{Rushchanskii2010,Rowan2010,Goian2011}.  Our samples, which have approximately 65\% ceramic density and were made in an identical way to those in Ref.~\cite{Sushkov2010}, can be partially polarized using moderate voltage ($\leq 3\kV$ or approximately $20$~kV/cm).  The magnetic Eu$^{2+}$ ions are responsible for paramagnetic behavior above approximately 1.9~K and behavior consistent with anti-ferromagnetism at lower temperatures~\cite{Rushchanskii2010}.   

\begin{figure}
 \center
 \psfrag{001}[lc][lc]{\footnotesize \shortstack{Connections to top \\ of cryostat}}
 \psfrag{002}[lc][lc]{\footnotesize \shortstack{Top sample \\ HV cable}}
 \psfrag{003}[lc][lc]{\footnotesize \shortstack{Superconducting \\ shield (inner)}}
 \psfrag{004}[lc][lc]{\footnotesize SQUID}
 \psfrag{005}[lc][lc]{\footnotesize \shortstack{Top sample ground \\ return line}}
 \psfrag{006}[lc][lc]{\footnotesize Top electrode}
 \psfrag{007}[lc][lc]{\footnotesize Top sample}
 \psfrag{008}[lc][lc]{\footnotesize Pickup loop}
 \psfrag{009}[lc][lc]{\footnotesize Ground plane}
 \psfrag{010}[lc][lc]{\footnotesize Bottom sample}
 \psfrag{016}[lc][lc]{\footnotesize Bottom electrode}
 \psfrag{012}[lc][lc]{\footnotesize \shortstack{Bottom sample \\ ground return}}
 \psfrag{013}[lc][lc]{\footnotesize \shortstack{Magnetic field \\ coils}}
 \psfrag{014}[lc][lc]{\footnotesize \shortstack{Bottom sample \\ HV cable}}
 \psfrag{015}[lc][lc]{\footnotesize \shortstack{Superconducting \\ shield (outer)}}

 \psfrag{020}[cr][cr]{\footnotesize $\hat{x}$}
 \psfrag{021}[tc][tc]{\footnotesize $\hat{y}$}
 \psfrag{022}[lc][lc]{\footnotesize $\hat{z}$}
 \includegraphics{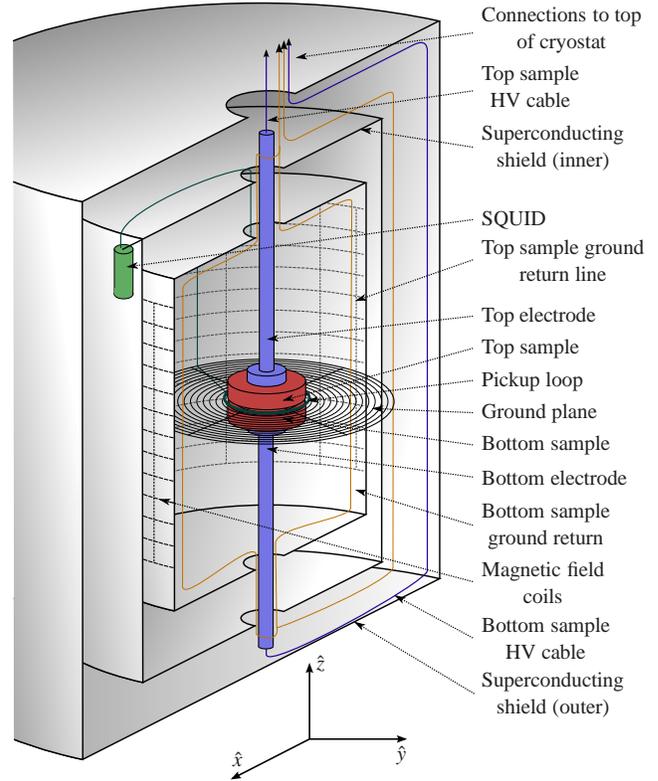}
 \caption{\label{fig:schematic} (Color online) A cut-through schematic of the $e$EDM experiment.  Note the coordinate system in the bottom of the figure.}
\end{figure}

The sample magnetization induced by the $e$EDM is given by
\begin{equation}
 \label{eq:edmmag}
 M = \frac{\chi_m d_e E^*}{\mu_a}\ ,
\end{equation}
where $\chi_m$ is the magnetic susceptibility, $d_e$ is the $e$EDM of the electron, $\mu_a$ is the magnetic moment of the Eu$^{2+}$ ion, and $E^*$ is the effective electric field.  As shown for a similar, dielectric material, Gd$_3$Ga$_5$O$_{12}$, the effective electric field is proportional to the displacement of the Eu$^{2+}$ with respect to the center of the oxygen octahedron around it~\cite{Mukhamedjanov2003}.  This displacement has been computed to be equal to half that of the displacement of the Ti$^{4+}$ ions with respect to the O$^{2-}$~\cite{Rushchanskii2010} and is therefore proportional to the polarization of the sample, i.e., $E^* = k P$.  Using this displacement and the results in Ref.~\cite{Mukhamedjanov2003}, we conservatively predict $k \approx (10\mbox{ MV/cm})/(1\mbox{ $\mu$C/cm$^2$})$.  The EDM interaction [Eq.~\ref{eq:edmmag}] can be viewed as a first order, linear magnetoelectric (ME) effect in the sample.  In this picture, the free energy of the sample $\tilde{\Phi}$ is modified by a linear term $\alpha' H P$, where $\alpha' = \chi_m d_e k/\mu_a$ and $H$ is the applied magnetic field.  Because the sample is cooled in a zero electric field and the experiment is operated at 4.2~K where the sample is paramagnetic, both parity and time symmetries are conserved in the crystal.  A non-zero $\alpha'$ can therefore only arise because of the $e$EDM~\cite{Smolenskii1982}.

A cut-through schematic of the experimental apparatus is shown in Fig.~\ref{fig:schematic}.  Two disc-shaped samples of diameter $12.6$~mm and height $1.7$~mm are held onto a centrally located ground plane by two electrodes.  Like most of the cryogenic components, the ground plane is constructed from G10 fiberglass but is coated with graphite to make the surface conductive.  An 8-turn superconducting Nb-Ti alloy pickup loop is wound inside the ground plane.  The pickup loop transfers the flux generated by the magnetization of the samples to a superconducting quantum interference device (SQUID) that is used as a magnetometer.  Because of the geometry of the samples, there are demagnetizing fields that lead to suppression of the magnetic flux detected by the SQUID~\cite{Sushkov2009}. To electrically polarize the samples, voltage is generated by a custom-built high voltage supply and applied via graphite-painted electrodes on the flat surfaces of the samples.  Additional leads from the ground planes attach to a high dynamic range, transimpedance amplifier~\cite{Eckel2012a}, with which currents that flow through the sample are measured.  The polarization is determined by numerically integrating the measured current.  Such numerical integration is accurate only to an arbitrary constant and thus measures the change in polarization but not the {\it absolute} polarization.

Two layers of superconducting magnetic shields made of 1~mm thick, 99.9\% pure Pb foil surround the sample region.  This shielding offers a minimum shielding factor of $10^8$ for time-varying magnetic fields.  However, during cooling of the experiment, they trap ambient magnetic fields as they undergo the superconducting transition.  This trapped field can be canceled using superconducting magnetic field coils wound on a cylindrical form of radius 5.2~cm and length 17~cm.  A solenoid coil applies a field $H_z$ parallel to the normal vector of the pickup loop (defined to be the $\hat{z}$ direction), and a cosine-$\theta$ type coil applies a field $H_x$ perpendicular to the normal vector of the pickup loop at a set azimuthal angle (defined to be the $\hat{x}$ direction).   Lastly, an anti-Helmholtz coil applies a magnetic field gradient $dH_z/dz$.

\begin{figure}
 %
 %
 \psfrag{000}[tc][tc]{$t$}%
 \psfrag{013}[cl][cl]{\small 40~kV/cm}%
 \psfrag{014}[cl][cl]{\small 100~$\mu$A}%
 \psfrag{015}[cl][cl]{\small 1~$\mu$C/cm$^2$}%
 \psfrag{016}[cl][cl]{\small 100~m$\Phi_0$}%
 \psfrag{017}[bc][bc]{\shortstack{SQUID\\Signal}}%
 \psfrag{018}[bc][bc]{Polarization}%
 \psfrag{019}[bc][bc]{Current}%
 \psfrag{020}[bc][bc]{\shortstack{Electric\\Field}}%

 \psfrag{001}[tr][tr]{\small $3\tau+t_s$}%
 \psfrag{002}[tr][tr]{\small $3\tau+t_p$}%
 \psfrag{003}[tr][tr]{\small $3\tau$}%
 \psfrag{004}[tr][tr]{\small $2\tau+t_s$}%
 \psfrag{005}[tr][tr]{\small $2\tau+t_p$}%
 \psfrag{006}[tr][tr]{\small $2\tau$}%
 \psfrag{007}[tr][tr]{\small $\tau+t_s$}%
 \psfrag{008}[tr][tr]{\small $\tau+t_p$}%
 \psfrag{009}[tr][tr]{\small $\tau$}%
 \psfrag{010}[tr][tr]{\small $t_s$}%
 \psfrag{011}[tr][tr]{\small $t_p$}%
 \psfrag{012}[tr][tr]{\small 0}%
 \includegraphics{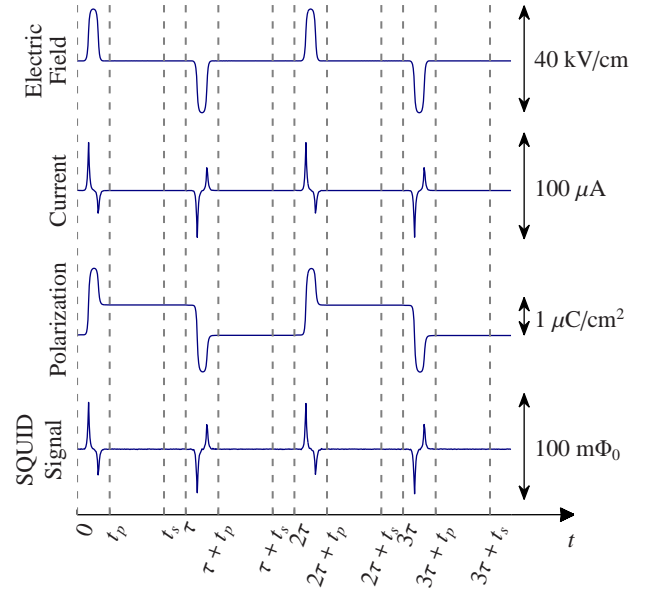}
 \caption{\label{fig:procedure} (Color online) The procedure for measuring the $e$EDM.  Electric field pulses (top) of duration $t_p$ are applied a time $\tau$ apart.  Each subsequent pulse reverses the polarization of the sample.  The current flow through the samples (second from the top) is numerically integrated to obtain the polarization of the samples (second from the bottom).  The SQUID signal (bottom) is averaged after each pulse between times $t_s$ and $\tau$, where $t_s$ is generally $0.8\tau$.  Shown on the right are typical orders of magnitude for the various applied fields and measured quantities.  If $d_e\approx 10^{-27}\ecm$, the size of the SQUID signal would be of order 1~n$\Phi_0$.}
\end{figure}

Fig.~\ref{fig:procedure} shows the experimental measurement procedure.  Electric field pulses separated by a time $\tau$ are applied to either the top sample, bottom sample, or both to modulate the remanent polarization.  Because Eq.~\ref{eq:edmmag} is linear in $P$, the $e$EDM-induced magnetization will be similarly modulated.  To measure the resulting modulation, the SQUID signal is averaged after allowing time for transients to settle.  To prevent background drifts in the signal from impacting the computation of the correlation, the average SQUID signals for four adjacent pulses in time are weighted by $\frac{1}{4}$, $-\frac{3}{4}$, $\frac{3}{4}$, $-\frac{1}{4}$ and summed.  This procedure determines the difference in the SQUID signal between the two polarization states $\Delta\Phi$ and eliminates the effect of a linear drift. $\Delta\Phi$ is then divided by the difference in the polarization $\Delta P$ to determine the correlation between the SQUID signal and the polarization.  This correlation $\Delta\Phi/\Delta P$ is proportional to the ME coefficient $\alpha'$ and thus $d_e$.

The predominant noise source is the SQUID magnetometer's intrinsic noise.  Above 1~Hz, the noise spectral density is approximately white at $3\ \mu\Phi_0/\sqrt{\text{Hz}}$.  Below 1~Hz, the noise of the SQUID rises roughly as $1/f$, where $f$ is the frequency.  Due to technical constraints, the fastest $\tau$ corresponds to a reversal frequency of $0.25\Hz$, within the $1/f$ noise regime of our SQUIDs.  Despite operating in the $1/f$ regime of the noise, the statistics for a data run are Gaussian.  Each data run comprises between 200 and 600 electric field pulses, and a Gaussian is fit to the distribution of $\Delta \Phi/\Delta P$.  The error of the best fit mean is used as the statistical error for that run.  The typical reduced $\chi^2$ for such a fit is near unity.  Because the samples are reversed at a frequency within the $1/f$ noise regime of the SQUIDs, the statistical errors of $\Delta \Phi/\Delta P$ tend to be an order of magnitude larger than those projected in Ref.~\cite{Sushkov2010}.

\begin{figure}
 \center
 %
 %
 \psfrag{025}[cc][cc]{Electric field (kV/cm)}%
 \psfrag{026}[bc][bc]{SQUID signal (m$\Phi_0$)}%
 \psfrag{027}[cc][cc]{Time (s)}%
 %
 %
 %
 \psfrag{015}[cc][cc]{\small $0$}%
 \psfrag{016}[cc][cc]{\small $2$}%
 \psfrag{017}[cc][cc]{\small $4$}%
 \psfrag{018}[cc][cc]{\small $6$}%
 \psfrag{019}[cc][cc]{\small $8$}%
 %
 %
 %
 \psfrag{000}[lc][lc]{\small }%
 \psfrag{001}[lc][lc]{\small -10}%
 \psfrag{002}[lc][lc]{\small 0}%
 \psfrag{003}[lc][lc]{\small 10}%
 \psfrag{004}[lc][lc]{\small 20}%
 \psfrag{005}[lc][lc]{\small -20}%
 \psfrag{006}[lc][lc]{\small -10}%
 \psfrag{007}[lc][lc]{\small 0}%
 \psfrag{008}[lc][lc]{\small 10}%
 \psfrag{009}[lc][lc]{\small }%
 \psfrag{010}[rc][rc]{\small }%
 \psfrag{011}[rc][rc]{\small 2}%
 \psfrag{012}[rc][rc]{\small 4}%
 \psfrag{013}[rc][rc]{\small 6}%
 \psfrag{014}[rc][rc]{\small 8}%
 \psfrag{020}[rc][rc]{\small 5}%
 \psfrag{021}[rc][rc]{\small 10}%
 \psfrag{022}[rc][rc]{\small 15}%
 \psfrag{023}[rc][rc]{\small 20}%
 \psfrag{024}[rc][rc]{\small }%
 %
 \includegraphics{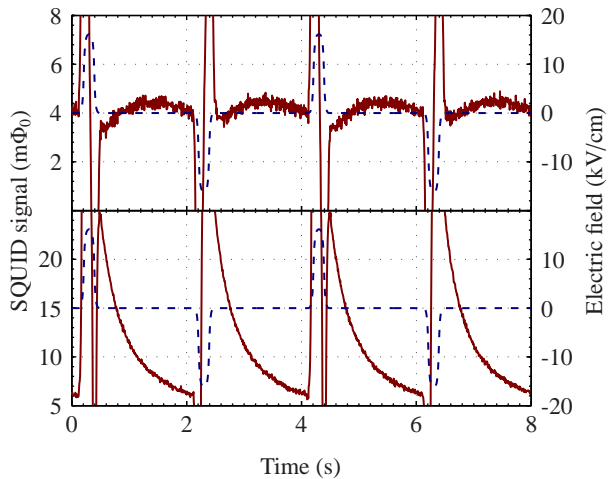}
 \caption{\label{fig:heating_example} (Color online) Example of the difference of the heating decay transient.  The blue dashed lines show the applied electric field pulses and the red solid lines show the resulting SQUID signal.  The large features in the SQUID signal seen during the electric field pulses are caused by the current that flows during the polarization reversal.  After the reversal, the heated sample returns to equilibrium with the LHe bath, which can be seen as the decay after the pulse.  These data were taken in the presence of an $H_x$ field (top panel) and an $H_z$ field (bottom panel), each approximately 1~mG.}
\end{figure}

Several systematic effects in the experiment can generate a non-zero $\Delta\Phi/\Delta P$ and can therefore mask or mimic the linear ME effect due to the $e$EDM.  For example, if the samples are in a non-zero magnetic field, a change in the temperature of the sample(s) will lead to a change in permeability that will subsequently change the flux through the SQUID.  Because of the dissipation inherent to ferroelectrics, polarization reversals heat the sample(s).  As the samples return to equilibrium with the liquid helium bath, a transient can be seen in the SQUID signal, as shown in Fig.~\ref{fig:heating_example}.  Provided this heating is equal when the sample polarization is switched from $+\hat{z}$ to $-\hat{z}$ (a negative pulse) and $-\hat{z}$ to $+\hat{z}$ (a positive pulse), the heating transients are identical for positive and negative remanent polarizations, and there is no systematic effect.  A measure of the amount of heat released by a given pulse can be derived from the integral of $P\cdot dE$, where $P$ is the polarization and $E$ is the applied electric field~\cite{JacksonEM}, and is of the order of 1~mJ per pulse.

To quantify the size of the resulting $\Delta\Phi/\Delta P$, magnetic fields were applied and the electric field pulses were deliberately unbalanced to produce different heating for positive and negative pulses.  The resulting correlation was measured in this manner for each reversal frequency and for each sample.  The correlations were then fit to $\Delta \Phi/\Delta P = a \Delta Q p$, where $\Delta Q$ is the difference in heat released between a positive and negative pulse, $p$ is a proxy for the magnetic field, and $a$ is a tunable constant.  As shown in Fig.~\ref{fig:heating_example}, the transient is significantly different for $H_x$ vs. $H_z$ fields; for this reason, the fits for the correlation use $p = \langle d\Phi/dt\rangle$ as a proxy for the strength of $H_z$ and $p = \langle d^2\Phi/dt^2\rangle$ as a proxy for the strength of $H_x$.  The resulting fits to experimental data confirm the validity of these proxies.  The best fit values for $a$ are used to predict the size of the correlation when the magnetic field is close to zero and the electric field pulses are symmetric.  In this configuration, it is not known {\it a priori} what type of field envelops the samples; therefore, the most likely correlation for both an $H_z$ field and an $H_x$ field is computed.  The resulting predictions are used as a 1-$\sigma$ systematic error without applying any correction.

In addition to this heating effect, the higher-order ME effect that is present in titanates can also generate a non-zero $\Delta \Phi/\Delta P$.  Given the symmetries present in our sample, the magnetization induced by the higher-order ME effect will be given by $M = \delta \chi_m P^2 H$, where $P$ is the {\it absolute} polarization.  Using the same experimental apparatus, the constant $\delta$ was measured for this material; details will be presented in a later paper in preparation.  Because the magnetoelectric-induced magnetization depends on $P^2$, a non-zero correlation will result only if the two different absolute polarization states in the modulation have different magnitudes.  Thus, the error in determining the absolute zero of polarization will determine the maximum possible difference in $P^2$ when the polarization is reversed.  The error in the absolute zero of $P$ is taken to be $0.1\mbox{ $\mu$C/cm$^2$}$ at 95\% confidence, which is motivated by the fidelity with which samples can be depolarized using electric fields.  Depolarization effectively resets the constant of integration in the determination of the polarization and thus the fidelity limits our knowledge of the absolute zero of the polarization.  Using this error estimate for the absolute measurement of $P$, a $\Delta\Phi/\Delta P$ is computed and used as a systematic error.

Because of the inherent dissipation present in ferroelectrics, the sample takes some time to reach the final polarization state after the electric field is applied.  This phenomenon is known as dielectric relaxation~\cite{Jonscher1999}.  As the sample relaxes to its final state, current continues to flow through the sample.  This current scales as $t^{-1}$, where $t$ is the time since the polarization reversal.  To suppress this dielectric relaxation, an additional time-varying voltage (maximum $40\V$) is applied using a proportional-integrator-differentiator (PID) circuit to force the net current to zero.  To estimate a $\Delta\Phi/\Delta P$ that may result, the SQUID response during the electric field pulse is used to calculate the sensitivity of the SQUID to the current through each sample.  The effect on the SQUID signal due to any current that is not suppressed by the PID is then computed and used to estimate the correlation.  The correlation due to dielectric relaxation is then taken to be a 1-$\sigma$ systematic error.

\begin{figure}
 \center
 %
 %
 \psfrag{019}[cc][cc]{$\alpha$ ($10^{-21}$ s/m)}%
 \psfrag{020}[cc][cc]{Run collection number}%
 \psfrag{021}[cl][cl]{$\overbrace{\text{~~~~~~~~~~~~~~~~~~~~~~~~~~~~~}}^\text{\small bottom sample}$}%
 \psfrag{022}[cl][cl]{$\overbrace{\text{~~~~~~~~~~~~~~~~~~~~}}^\text{\small top sample}$}%
 \psfrag{023}[cl][cl]{$\overbrace{\text{~~~~~~~~~~~~~~~~~~~~}}^\text{\small both samples}$}%
 \psfrag{024}[bc][bc]{Best fit $d_e$ ($10^{-25}\ e$cm)}%
 %
 %
 %
 \psfrag{007}[cc][cc]{\small $0$}%
 \psfrag{008}[cc][cc]{\small $2$}%
 \psfrag{009}[cc][cc]{\small $4$}%
 \psfrag{010}[cc][cc]{\small $6$}%
 \psfrag{011}[cc][cc]{\small $8$}%
 %
 %
 %
 \psfrag{012}[rc][rc]{\small $-30$}%
 \psfrag{013}[rc][rc]{\small $-20$}%
 \psfrag{014}[rc][rc]{\small $-10$}%
 \psfrag{015}[rc][rc]{\small $0$}%
 \psfrag{016}[rc][rc]{\small $10$}%
 \psfrag{017}[rc][rc]{\small $20$}%
 \psfrag{018}[rc][rc]{\small $30$}%
 \psfrag{000}[lc][lc]{\small $-15$}%
 \psfrag{001}[lc][lc]{\small $-10$}%
 \psfrag{002}[lc][lc]{\small $-5$}%
 \psfrag{003}[lc][lc]{\small $0$}%
 \psfrag{004}[lc][lc]{\small $5$}%
 \psfrag{005}[lc][lc]{\small $10$}%
 \psfrag{006}[lc][lc]{\small $15$}%
 %
 %
 \includegraphics{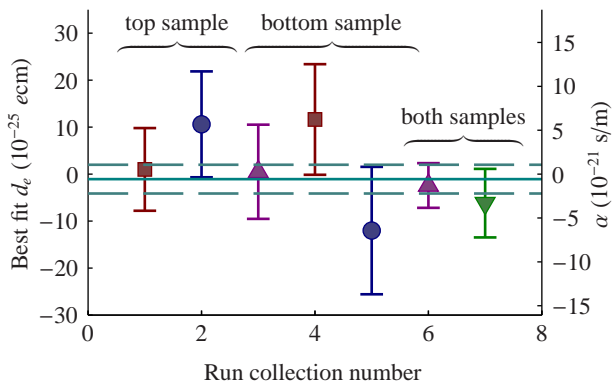}
 \caption{\label{fig:edm_results} (Color online) Final best fit $e$EDM values with statistical error bars.  Data were recorded at reversal frequencies of 0.25~Hz (violet, upward triangles), 0.167~Hz (red squares), 0.125~Hz (green, downward triangles), and 0.1~Hz (blue circles).  The annotations show which sample(s) were driven.  The solid cyan line shows the best fit mean, and the dashed cyan lines show its 1-$\sigma$ statistical error.}
\end{figure}

\begin{table}
 \center
 \begin{tabular}{lrrr}
 & Top	& Bottom & Both\\
 \hline
 Heating ($H_x$)	& $0.71$	& $1.82$	& $-0.07$ \\
 Heating ($H_z$)	& $-0.05$	& $-0.72$	& $-0.02$ \\
 Dielectric relaxation	& $-0.97$	& $-0.11$	& $0.70$ \\
 Higher-order ME effect	& $1.40$	& $0.26$	& $0.47$ \\
 \hline\hline
 \end{tabular}
 \caption{\label{tab:sys_breakdown} Breakdown of the systematic errors in the experiment by source and which sample(s) were driven.  Units of the table are $10^{-25}\ecm$.}
\end{table}

The total integrated time for the data used in the final analysis is approximately 1 hour and 40 minutes.  All data where the same sample(s) are driven at the same reversal frequency and with the same amplitude electric field pulses were averaged together, weighted by their statistical errors.  The correlation is then converted into a linear ME coeffecient and an equivalent $d_e$.  To enable comparision with linear ME coefficients that are expressed in units of s m$^{-1}$, we define $\alpha = \chi_e\epsilon_0\alpha'$, where $\chi_e = P/\epsilon_0E\approx 700$ is an effective electrical susceptibility for the ferroelectric.  The results are shown in Fig.~\ref{fig:edm_results} and are consistent with zero.

A breakdown of the systematics is shown in Tab.~\ref{tab:sys_breakdown}.  The higher-order ME effect produces a significant systematic effect because of the conservative estimate of our knowledge of the absolute polarization.  The systematic due to the heating shows complicated behavior, and is significantly less when both samples are driven.  The reason for this reduction is twofold.  When driving both samples, there is a significant rejection of the effect of a transverse field because $H_x$ couples to the two samples in the opposite way.  Second, the asymmetry in the heating when both samples were used was measured to be nearly equal and opposite, leading to rejection of $H_z$.

The final best fit results are $d_e = (-1.07\pm3.06_\text{stat}\pm 1.74_\text{sys})\times10^{-25}\ecm$ and  $\alpha = (-0.57\pm1.64_\text{stat}\pm 0.93_\text{sys})\times10^{-21}\mbox{ s/m}$~\footnote{The error in $d_e$ does not include any uncertainty from the theoretical estimate of $E^*$.}.  This result implies an $e$EDM limit of $|d_e|<6.05\times10^{-25}\ecm$ (90\% confidence).  Compared to previous solid state eEDM measurements, this limit is approximately a factor of ten improvement over Ref.~\cite{Heidenreich2005} and a factor of three better than Ref.~\cite{Kim2011}.

In conclusion, we have built and operated an experiment that has established an upper limit on the $e$EDM better than any previously published solid-state experiment.  The typical remanent polarization of Eu$_{0.5}$Ba$_{0.5}$TiO$_3$ of $0.5\uCcm$ offers a large effective electric field that interacts with the EDM, approximately 700 times larger than that obtained in Ref.~\cite{Kim2011} where dielectric Gd$_3$Ga$_5$O$_{12}$ was used.  The ultimate EDM limit can be improved in future versions of the experiment by identifying and suppressing the sources of excess noise in the SQUID magnetometers below $1\Hz$.   Further suppression of systematics, such as heating and dielectric relaxation, could be obtained by improving magnetic shielding and optimizing the current feedback system.  Alternatively, these systematics may be suppressed by using either a low-loss ferroelectric (e.g., Eu$_{0.5}$Ba$_{0.25}$Sr$_{0.25}$TiO$_3$~\cite{Goian2012}) or paraelectric (e.g., SrTiO$_3$ doped with Eu$^{2+}$~\cite{Muller1979,Viana1994}).

The authors would like to thank N.A. Spaldin, O.P. Sushkov, D. Budker, L.S. Bouchard and D.P. DeMille for useful discussions.  We would also like to thank J. Haase, N. Georgieva, S. Kamba for providing preliminary measurements of Eu$_{0.5}$Ba$_{0.5}$TiO$_3$.  This work was supported by Yale University.

\end{document}